\documentclass[aps,prl,twocolumn]{revtex4}

\usepackage{graphicx}
\usepackage{amsmath,amssymb}
\newcommand{\Sr}{\ensuremath{^{87}}Sr}

\begin{document}

\title{Comparison of Two Independent Sr Optical Clocks with $1\times10^{-17}$ Stability at 10$^3$~s}
\author{T.L. Nicholson, M.J. Martin, J.R. Williams, B.J. Bloom, M. Bishof}
\affiliation{JILA, National Institute of Standards and Technology and University of Colorado,
Department of Physics, University of Colorado, Boulder, Colorado 80309-0440, USA}
\author{M.D. Swallows}
\altaffiliation{Present address: AOSense, Sunnyvale, CA}
\affiliation{JILA, National Institute of Standards and Technology and University of Colorado,
Department of Physics, University of Colorado, Boulder, Colorado 80309-0440, USA}
\author{S.L. Campbell, and J. Ye}
\affiliation{JILA, National Institute of Standards and Technology and University of Colorado,
Department of Physics, University of Colorado, Boulder, Colorado 80309-0440, USA}

\date{\today}

\begin{abstract}
Many-particle optical lattice clocks have the potential for unprecedented measurement precision and stability due to their low quantum projection noise. However, this potential has so far never been realized because clock stability has been limited by frequency noise of optical local oscillators. By synchronously probing two $^{87} \mbox{Sr}$ lattice systems using a laser with a thermal noise floor of $1 \times 10^{-15}$, we remove classically correlated laser noise from the intercomparison, but this does not demonstrate independent clock performance. With an improved optical oscillator that has a $1 \times 10^{-16}$ thermal noise floor, we demonstrate an order of magnitude improvement over the best reported stability of any independent clock, achieving a fractional instability of $1 \times 10^{-17}$ in 1000~s of averaging time for synchronous or asynchronous comparisons. This result is within a factor of 2 of the combined quantum projection noise limit for a 160 ms probe time with $\sim$10$^3$ atoms in each clock.  We further demonstrate that even at this high precision, the overall systematic uncertainty of our clock is not limited by atomic interactions. For the second Sr clock, which has a cavity-enhanced lattice, the atomic-density-dependent frequency shift is evaluated to be $-3.11 \times 10^{-17}$ with an uncertainty of $8.2 \times 10^{-19}$.
\end{abstract}

\maketitle

Precise time keeping is foundational to technologies such as high-speed data transmission and communication, GPS and space navigation, and new measurement approaches for fundamental science. Given the increasing demand for better synchronization, more precise and accurate clocks are needed, motivating the active development of atomic clocks based on optical transitions. Several optical clocks have surpassed the systematic uncertainty of the primary Cs standard~\cite{Bize2005,Parker2010}. Two examples are the NIST trapped $\mbox{Al} ^{+}$ single ion clock, with a systematic uncertainty of $8.6 \times 10^{-18}$~\cite{Chou2010}, and the JILA $^{87} \mbox{Sr}$ neutral atom lattice clock, at the $1.4 \times 10^{-16}$ level~\cite{Ludlow2008,Swallows2012}. The field of optical atomic clocks has been very active in recent years, with many breakthrough results coming from both the ion clock \cite{Oskay2006,Chou2010,Huntemann2012,King2012,Dube2005} and lattice clock \cite{Swallows2011,Westergaard2011,Takamoto2011,Middelmann2012,Yamaguchi2012,Schiller2012} communities.

In principle, the stability of an optical lattice clock can surpass that of a single-ion standard because the simultaneous interrogation of many neutral atoms reduces the quantum projection noise (QPN) of the lattice clock~\cite{Itano1993,Santarelli1999}. QPN determines the standard quantum limit to the clock stability, and it can be expressed as

\begin{equation}
\sigma_{\mathrm{QPN}}(\tau) = \frac{\chi}{\pi Q} \sqrt{\frac{T_{c}}{N \tau}}.
\end{equation}

\noindent Here, $\sigma_{\mathrm{QPN}}(\tau)$ is the QPN-limited fractional instability of a clock, $Q$ is the quality factor of the clock transition, $N$ is the number of atoms, $T_{c}$ is the clock cycle time, $\tau$ is the averaging time (in seconds), and $\chi$ is a numerical factor near unity that is determined by the line shape of the clock transition spectroscopy. In a typical lattice clock, $N$ is on the order of $10^{3}$. In the case of the $\mbox{Al}^{+}$ ion clock, $N=1$, and a fractional instability of $2.8 \times 10^{-15} / \sqrt{\tau}$ for a two-clock comparison has been demonstrated~\cite{Chou2010}. For typical values of $T_{c}$ and $N$, a QPN-limited \Sr\ lattice clock could potentially reach a given stability 500 times faster than the $\mbox{Al} ^{+}$ clock.

Despite this promise, thus far the instability of lattice clocks has been far worse than the QPN limit. Instead, demonstrated lattice clock instability has been dominated by downsampled broadband laser noise (the Dick effect~\cite{Santarelli1998}) at a few times $10^{-15} / \sqrt{\tau}$, similar to that of the best ion systems~\cite{Ludlow2008,Lodewyck2009,Westergaard2011,Takamoto2011}. To improve the precision of lattice clock systematic evaluations while avoiding the challenge of building more stable clock lasers, a synchronous interrogation method can be implemented~\cite{Bize2000,Takamoto2011}. Synchronous interrogation facilitates laser-noise-free differential measurements between two atomic systems; however, in this approach, these systems are not independent clocks.

In this work, we achieve instability at the $10^{-16}/\sqrt{\tau}$ level for two independent \Sr\ optical lattice clocks. Using a new clock laser stabilized to a 40~cm optical reference cavity~\cite{Swallows2012} with a thermal noise floor~\cite{Notcutt2006} of $1 \times 10^{-16}$, we directly compare two independently operated \Sr\ clocks. The combined stability of these clocks is within a factor of 2 of the QPN limit, reaching $1 \times 10^{-17}$ stability in only 1000~s. We also use synchronous interrogation to study the effect of laser noise on clock stability, demonstrating its effectiveness in removing correlated noise arising from a 7~cm cavity with a $1 \times 10^{-15}$ thermal noise floor. Operating with the 40~cm cavity, on the other hand, synchronous and asynchronous interrogations (the latter of which demonstrates independent clock performance) yield nearly the same measurement precision for a given averaging time.

This high measurement precision will permit much shorter averaging times for a range of applications, including investigations of systematic uncertainties in lattice clocks. In particular, we are able to characterize one of the most challenging systematics in a many-particle clock---the density-related frequency shift~\cite{Campbell2009,Swallows2011,Lemke2011,Ludlow2011}---at an uncertainty below $1 \times 10^{-18}$ for our second Sr clock. The only remaining major systematic uncertainty for lattice clocks is the blackbody-radiation-induced Stark shift~\cite{Porsev2006,Campbell2008,Sherman2012,Middelmann2012}. One can mitigate this effect by trapping atoms in a well-characterized blackbody environment or cold enclosure~\cite{Middelmann2011}.

Our previous clock comparisons involved referencing our first generation \Sr\ clock (Sr1) to various clocks at NIST using a 3.5~km underground fiber optic link~\cite{Ludlow2008,Campbell2008}. To evaluate the stability of the \Sr\ clock at the highest possible level, we constructed a second Sr clock (Sr2) for a direct comparison between two systems with similar performance. In both systems, \Sr\ atoms are first cooled with a Zeeman slower and a magneto-optical trap (MOT) on a strong 30~MHz transition at 461~nm. Then a second MOT stage, operating on a 7.5~kHz intercombination transition at 689~nm, cools the atoms to a few $\mu$K. Atoms are then loaded from their 689~nm MOTs into 1D optical lattices and are nuclear spin polarized. The lattices operate near the ``magic" wavelength at 813~nm where the differential AC Stark shift for the $^{1}S_{0}$ and $^{3}P_{0}$ clock states is identically zero~\cite{Ye2008}.

The lattice for Sr1 is made from the standing wave component of a retroreflected optical beam focused to a 32~$\mu \mbox{m}$ radius. The power in one direction of this beam is 140~mW, corresponding to measured trap frequencies of 80~kHz along the lattice axis and 450~Hz in the radial direction. From this trap frequency we estimate a 22~$\mu \mbox{K}$ trap depth. The Sr2 lattice utilizes an optical buildup cavity so that laser power in one direction of this lattice is 6~W. The cavity has a finesse of 120 and is mounted outside our vacuum chamber. The intracavity beam radius for this lattice is 160~$\mu \mbox{m}$, which yields a much greater trap volume. Trap frequencies in this lattice are 100~kHz and 120~Hz in the axial and radial directions, respectively. We estimate a 35~$\mu \mbox{K}$ trap depth for the cavity-enhanced lattice.

\begin{figure}
  \centering
  \includegraphics[width=\linewidth]{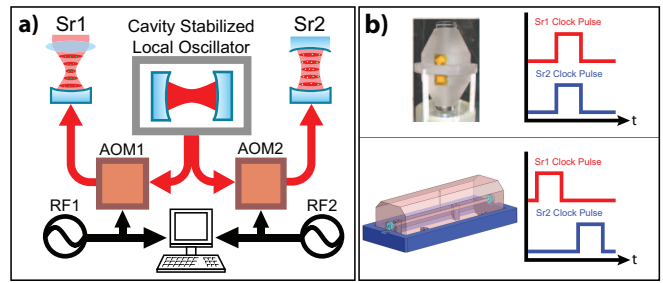}
  \caption{(a) A cavity-stabilized diode laser is split and sent to each of the lattice clocks. To ensure that both clock laser beams have independent frequency control, Sr1 and Sr2 have separate AOMs. The two clocks have different lattice geometries: Sr1 uses a 1D retroreflected lattice and Sr2 uses a 1D cavity-enhanced lattice. The independent clock laser beams are locked to the ${}^{1}S_{0} \rightarrow {}^{3}P_{0}$ transition by feeding the measured clock transition frequency back to the rf frequencies of the AOMs. The rf frequencies are recorded to determine the difference between the two clocks. (b) Clock comparisons using our 7~cm vertical cavity-stabilized laser (top) required synchronizing the clock probe pulses to perform correlated spectroscopy. Clock comparisons with our lower noise 40~cm horizontal cavity (bottom) used asynchronous pulses to ensure independent clock operation. Synchronous measurements with the 40~cm cavity (not depicted) were also performed.}
  \label{fig:laser_systems}
\end{figure}

The optical local oscillator for the Sr1 and Sr2 systems is derived from a common cavity-stabilized diode laser at 698~nm, but two different acousto-optic modulators (AOMs) provide independent optical frequency control for each system [Fig.~\ref{fig:laser_systems}(a)]. For all measurements presented in this Letter, we use Rabi spectroscopy with a 160~ms probe time, corresponding to a Fourier-limited linewidth of 5~Hz. For the stability measurements, we use 1000 atoms for Sr1 and 2000 atoms for Sr2.  The optical frequency is locked to the clock transition using a digital servo that provides a correction to the AOM frequency for the corresponding clock.

\begin{figure}
  \centering
  \includegraphics[width=\linewidth]{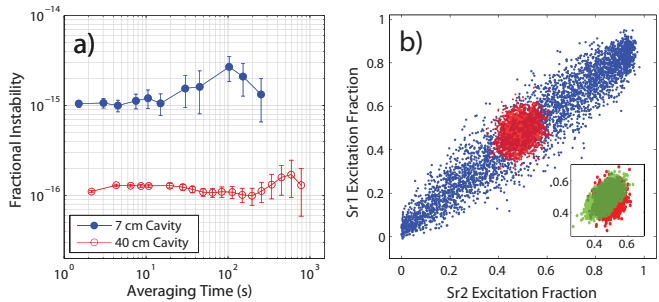}
  \caption{(a) The measured thermal noise floor of the two optical reference cavities. The stability of the 7~cm cavity (closed circles) was measured by comparing two cavities of the same design. For the 40~cm cavity (open circles), we determine its frequency stability from a measurement based on the atomic reference. We lock this laser to the \Sr\ clock transition and subtract off a residual cavity drift of $\sim$1.4 mHz/s. These data include contributions from other technical noise and thus represents an upper bound on the thermal noise floor. (b) A scatter plot of the measured excitation fraction when the clock lasers are locked to the two Sr references. Each point represents the measured excitation fraction for Sr1 versus Sr2 for the same duty cycle. The blue points represent data taken under synchronous interrogation using the 7~cm reference cavity, showing a clear correlation arising from common-mode laser noise. The red points represent data taken under asynchronous interrogation with the low-noise 40~cm reference cavity, clearly indicating a lack of classical correlations. Instead, the distribution indicates near-QPN-limited performance for independent Sr1 and Sr2. The inset compares synchronous measurements using the 40~cm cavity (in green) with the asynchronous data using the same cavity. This distribution shows a slight correlation, indicating a small amount of residual laser noise.}
  \label{fig:thermal_noise}
\end{figure}

To provide a quantitative understanding of the role of laser noise in our clock operations, we use two different clock lasers in our experiment. The first clock laser is frequency stabilized to a vertically oriented 7~cm long cavity with a thermal noise floor of $1 \times 10^{-15}$ ~\cite{Ludlow2007}. This 7~cm reference cavity was used in much of our previous clock work and represented the state-of-the-art in stable lasers until recently. The second laser is stabilized to a horizontal 40~cm long cavity with a thermal noise floor of $1 \times 10^{-16}$ [Fig.~\ref{fig:thermal_noise}(a)], which is similar to the record performance achieved with a silicon-crystal cavity~\cite{Kessler2012}. The greater cavity length and use of fused silica mirror substrates both reduce the thermal noise floor of this laser~\cite{Swallows2012}. Other significant improvements for the 40~cm system include a better vacuum, active vibration damping, enhanced thermal isolation and temperature control, and an improved acoustic shield.

When comparing the two \Sr\ systems using the 7~cm reference cavity, the probe pulses for the Sr1 and Sr2 clock transitions are precisely synchronized [Fig.~\ref{fig:laser_systems}(b)]. The responses of both digital atomic servos are also matched. This synchronous interrogation allows each clock to sample the same laser noise; therefore, the difference between the measured clock transition frequencies for Sr1 and Sr2 benefits from a common-mode rejection of the laser noise. Because of this common-mode laser noise, simultaneous measurements of the excitation fraction for the Sr1 and Sr2 atomic servos show classical correlations [Fig.~\ref{fig:thermal_noise}(b)], as evidenced by the distribution of these measurements in the shape of an ellipse stretched along the correlated (diagonal) direction. The minor axis of this distribution indicates uncorrelated noise such as QPN.

The 40~cm cavity supports a tenfold improvement in laser stability, and we estimate that the Dick effect contribution is close to that of QPN for clock operation. To test this, we operate the two clocks asynchronously, where the clock probes are timed such that the falling edge of the Sr1 pulse and the rising edge of the Sr2 pulse are always separated by 10~ms [Fig.~\ref{fig:laser_systems}(b)]. During this asynchronous comparison, the two clocks sample different laser noise, preventing common-mode laser noise rejection. The Sr1--Sr2 excitation fraction scatter plot [Fig.~\ref{fig:thermal_noise}(b)] resembles a 2D Gaussian distribution, which is consistent with both clocks being dominated by uncorrelated white noise. Synchronous comparisons with the 40~cm cavity were also performed, indicating a similar distribution for the scatter plot of the Sr1 vs. Sr2 excitation [Fig.~\ref{fig:thermal_noise}(b) inset].

\begin{figure}
  \centering
  \includegraphics[width=\linewidth]{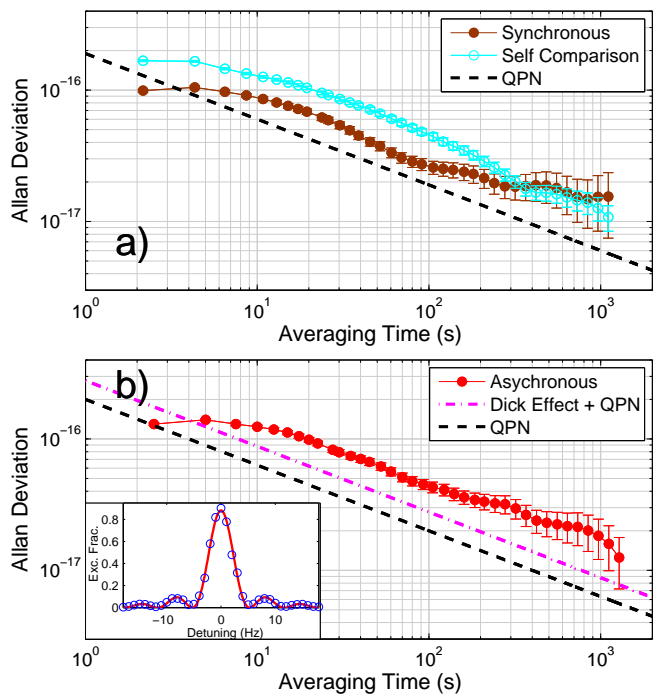}
  \caption{(a) The Allan deviation of a synchronous comparison (closed circles) between Sr1 and Sr2 with the low-noise 40~cm cavity. The self-comparison (open circles) is $(\nu_{1} - \nu_{2})/\sqrt{2}$, where $\nu_{1}$ and $\nu_{2}$ are the frequencies to which the two servos are locked. Dividing $(\nu_{1} - \nu_{2})$ by $\sqrt{2}$ extrapolates the self-comparison stability to the expected performance of a comparison between the Sr2 system and an identical clock. The dashed line indicates the QPN limit. (b) An asynchronous comparison between the two Sr clocks (also taken with the 40~cm cavity). The Allan deviation of the comparison fits to $4.4 \times 10^{-16} / \sqrt{\tau}$. The estimated Dick effect is roughly equal to the predicted QPN of $2.0 \times 10^{-16} / \sqrt{\tau}$ (dashed line). The inset depicts typical scans of the clock transition (open circles). The red line is a fit to the data using the Rabi model. All stability data shown in this work represent the combined stability of the two systems. To infer a single clock stability, one would need to divide all the data by $\sqrt{2}$.}
  \label{fig:comparison}
\end{figure}

With this understanding of laser noise effects in our clocks, we now evaluate the clock stability. In the short term ($<$100 s) the clock stability is limited by laser noise and QPN, and in the long term ($\sim$1000~s) it is limited by drifting systematic shifts. Using the 40~cm cavity, we measure the short- and long-term stability in two ways. The first approach combines information from both a self-comparison and a synchronous comparison to infer the full stability of our clocks [Fig.~\ref{fig:comparison}(a)]. A self-comparison involves comparing two independent atomic servos on the Sr2 system~\cite{Swallows2011}. Updates for these two digital servos alternate for each experimental cycle. Thus the difference between these servo frequencies is sensitive to the Dick effect and QPN and therefore represents the short-term stability of an independent clock~\cite{Jiang2011,Hagemann2012}; however, it does not measure the clock's long-term stability as it is insensitive to all drifts at time scales greater than 5~s. The other component of this approach, the synchronous comparison, is sensitive to long-term drifts on either system, but in the short term it is free of correlated laser noise. Together these two data sets provide a complete picture of our clock's short- and long-term stability, and the small difference between them after about 10~s implies that our clocks are only minimally affected by correlated noise.

In the second approach, we measured the full stability of our clock with an asynchronous comparison, which is sensitive to both short- and long-term instability. Beyond the atomic servo response time ($>$20 s), an asynchronous comparison reflects the performance of two independent clocks. Analysis of the Dick effect for our asynchronous pulse sequence (and a thermal-noise-limited local oscillator) shows that our asynchronous comparison reproduces independent clock performance within 6\%. The Allan deviation of the comparison signal is shown in Fig.~\ref{fig:comparison}(b). These results demonstrate that one or both of our \Sr\ clocks reaches the $1 \times 10^{-17}$ level in 1000~s, representing the highest stability for an individual clock and marking the first demonstration of a comparison between independent neutral-atom optical clocks with a stability well beyond that of ion systems.

The enhanced stability of many-particle clocks can come at the price of higher systematic uncertainty due to density-dependent frequency shifts, which arise from atomic interactions. This shift has received a great deal of attention in recent years, with experiments and theory centered around schemes for explaining and minimizing this effect for optical lattice clocks~\cite{Swallows2011,Gibble2009,Rey2009,Lemke2011,Yu2010}. To operate at lower densities, the Sr2 system employs a large volume optical lattice created by a buildup cavity that results in a lower density shift than Sr1 \cite{Westergaard2011}. The large lattice volume also allows Sr2 to trap as many as 50~000 atoms under typical experimental conditions.

\begin{figure}
  \centering
  \includegraphics[width=\linewidth]{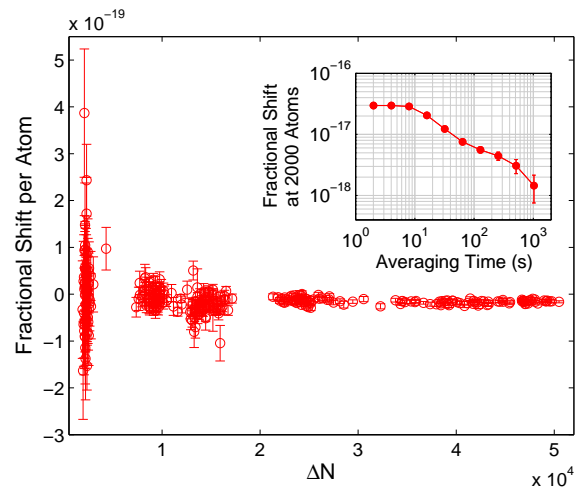}
  \caption{The measured Sr2 density shift as a function of $\Delta N$. Each point on this plot represents an average over a bin of 30 measurements of $(\nu_{\mathrm{high}} - \nu_{\mathrm{low}})/(N_{\mathrm{high}} - N_{\mathrm{low}})$. Our statistics show that the density shift for our trapping conditions is linear within our quoted uncertainty of $8.2 \times 10^{-19}$. \textbf{Inset}: A single 2000~s long density shift measurement with $\Delta N \simeq 41 000$. The shift per atom was measured and then scaled up to 2000 atoms for a typical running condition. This measurement shows that a single density shift evaluation for 1000~s using a large atom number modulation is sufficient for a $1 \times 10^{-18}$ clock.}
  \label{fig:density}
\end{figure}

We measure the Sr2 density shift with Rabi spectroscopy and synchronous interrogation, using our 7~cm cavity. In this case where laser noise dominates the single-clock instability, synchronous interrogation allows us to evaluate this systematic an order of magnitude faster than we could without the Sr1 reference. This measurement alternates between two independent atomic servos, one addressing a high atom number sample, $N_{\mathrm{high}}$, and one addressing a low atom number $N_{\mathrm{low}}$. The first (second) servo measures a frequency $\nu_{\mathrm{high}}$ ($\nu_{\mathrm{low}}$), and the corresponding $N_{\mathrm{high}}$ ($N_{\mathrm{low}}$) is recorded during each cycle. For a frequency shift that is linear in density, the quantity $(\nu_{\mathrm{high}} - \nu_{\mathrm{low}})/(N_{\mathrm{high}} - N_{\mathrm{low}})$ is the slope of the shift. For our greatest modulation amplitude of $\Delta N = N_{\mathrm{high}} - N_{\mathrm{low}} \simeq 47 000$, we determine that the uncertainty in the density shift per 2000 atoms (corresponding to an average density of 2 to 3 $\times ~ 10^{9} ~\mbox{cm}^{-3}$) reaches the $1 \times 10^{-18}$ level in 1000~s [Fig.~\ref{fig:density} inset].

To verify that the shift is linear in atom number, we vary $\Delta N$ by changing $N_{\mathrm{high}}$ while setting $N_{\mathrm{low}}$ to 2000--3000 atoms [Fig.~\ref{fig:density}]. We analyze the density shift data using the statistical analysis from our previous work~\cite{Swallows2011}. Our error bars are inflated by the square root of the reduced chi-square statistic $\chi_{\mathrm{red}}^{2}$ calculated for a model in which the density shift is directly proportional to our atom number. For this measurement, $\sqrt{\chi_{\mathrm{red}}^{2}} = 1.3$. The $\chi_{\mathrm{red}}^{2}$ statistic can differ from unity due to drifts in the calibration of the fluorescence signal used to measure our atom number, slight variations in the optical trapping conditions, or departures from a proportional model. We determine the Sr2 density shift of $(-3.11 \pm 0.08) \times 10^{-17}$ at 2000 atoms. At this atom number, the total shift is sufficiently small such that our clock is stable at the $1 \times 10^{-18}$ level in the presence of typical atom number drifts.

In summary, we have demonstrated comparisons between two independent optical lattice clocks with a combined instability of $4.4 \times 10^{-16} / \sqrt{\tau}$, with a single clock demonstrating $1 \times 10^{-17}$ level instability at 1000~s. We have also determined the density-dependent frequency shift uncertainty in our cavity-enhanced lattice at $8.2 \times 10^{-19}$, with single measurements averaging down to the $1 \times 10^{-18}$ level in 1000~s.

We acknowledge funding support from NIST, NSF, and DARPA. JRW is supported by NRC RAP. MB acknowledges support from NDSEG. SLC acknowledges support from NSF. We thank X. Zhang and W. Zhang for technical contributions.

\end{document}